\documentstyle[12pt]{article} 
\begin{document}
\topmargin-1.0cm
\leftmargin-1.0cm
\hsize=6.0truein
\vsize=10.0truein
\parskip=0.0pt
\parindent=15pt
\baselineskip=10pt
\begin{titlepage}
\begin{center}
{\huge The  random walks \vskip.2cm of \vskip.2cm  a Schwarzschild
 black hole} 
\end{center}
 
\vspace{5 mm}
 
\begin{center}
Marcelo Schiffer\footnote{On  leave from University  of Campinas\\e.mail
marcelos@astro.huji.ac.il}\\ Racah Institute of Physics\\
The Hebrew  Univerity of Jerusalem\\
Givat Ram, Jerusalem,91904\\
Israel
\end{center}
 
\vspace{10 mm}
 
\begin{center}
{\large Abstract}
\end{center}
 
The purpose of this paper is to show  that as a matter of principle, it is not
appropriate to consider   Schwarzschild black holes in  thermal
equilibrium   with  radiation because, even spinless and neutral holes undergo
fluctuations of charge and angular momentum. Therefore, there will be a spread 
of
these quantities around their zero-mean values. We  calculate these  spreads
for a black   hole in thermal contact with charged scalar particles and show
that angular momentum fluctuations are governed by the size of the cavity,
larger cavities yielding larger angular momentum fluctuations. This behaviour is
expected if  black hole  angular momentum fluctuations stem  from the  random
absorption and emission of quanta with  random  angular momenta from the
thermal bath.  Furthermore, in  the limit $R/2M \gg 1$  charge fluctuations
$\Delta Q^2/(c\hbar) \sim 1/(4\pi)$, that is to say,  they become scale
independent. This  is expected if the underlying physics of these fluctuations
is the  random absorption and emission of charged  quanta from the thermal bath.
The  independence of these fluctuations upon  the elementary charge of the
field is puzzling because it tells us  that either the scale of the elementary 
charge is fixed  by black hole physics to be   $\alpha \sim 1/4\pi$
(this gives an elementary  charge  which is only  three times the charge of the
electron), or the  underlying physical process responsible these fluctuations
is  not known  and remains to  be cleared up.

\vskip2.0cm
PACS: 04.70Dy, 04.62.+v
\end{titlepage}
\newpage

\section{Introduction}

The fact that black holes are subjected to the same thermodynamical laws as are
ordinary  systems is one of the most  intriguing findings and is likely  to be 
our best window toward an  elusive theory of quantum gravity. 
Therefore, exploring and exploiting this paradigm could  yield important clues
of the form  and  content of this still unveilled theory.

Thermodynamic equilibrium  between black holes and radiation is of
considerable interest and was matter  of investigation in the
past\cite{dav77}-\cite{hut}. In particular for
Schwarzschild holes, if was concluded that the  vessel  confining the
radiation  and black hole   cannot  be  larger  than some critical value
because  larger volumes  render the (total) entropy  a concave function
rather than convex. Although these studies are very  important for the 
understanding of  the conditions  of equilibrium of Schwarzschild black 
holes and radiation, they  are  incomplete in the sense that leave out an
important ingredient, namely that it is not possible  to prevent a
neutral and non-rotating black hole from  undergoing statistical
fluctuations of charge and angular momentum around their zero-mean 
values. Put in another way, these fluctuations,  although preserving the 
zero mean values of  charge  and angular momentum,  produce  mean  square 
deviations  of  these quantites. 

In this paper it is  shown that these thermodynamical fluctuations do not result
in  any new instability, regardless what the external parameters are.
Furthermore,  we estimate both  angular momentum and charge standard deviations.

This   is done by first assuming  a non-vanishing mean charge and mean angular
momentum and by taking at the very end the limit were they do vanish. The paper
is organized in the following way: in the next section we estimate angular
momentum fluctuations and  in the sequel charge fluctuations. The
last section is devoted to the discussion of possible implications.

\section{angular momentum fluctuations}

Consider a black hole of  angular momentum $J$ and mass $M$, in  equilibrium 
with thermal radiation  of massless scalar particles, inside a spherical 
vessel of radius $R$. The field operator  associated to  these   quanta
satisfies a Klein-Gordon  equation, which  is to be supplemented by a 
Dirichlet type  boundary  condition.  For this reason
\begin{equation}
\Phi (\vec{x},t) \approx j_l(\epsilon r) Y_{lm}(\theta,\phi) e^{-i \omega t} \,
. \end{equation}
were   $j_l(x)$  and $Y_{lm}(\theta,\phi)$   are  the spherical  Bessel
and harmonic functions, respectively. The field spectrum  is
provided by  the roots $x_{l,n}=\epsilon_{l,n} R$ of the spherical Bessel
functions $j_l(x_{l,n})=0$.

Assuming that the  system  has no net angular momentum, conservation of energy
and angular momentum  requires that  
\begin{eqnarray}
0&=&j(\Omega, T) + J  \nonumber \\ 
E&=& e(\Omega, T) + M \, 
\label{constraints}
\end{eqnarray}
where $j(\Omega, T)$ and $e(\Omega, T)$ are the  field's  mean  angular momentum
and mean energy. More precisely,
\begin{equation} 
j(\Omega, T) =
\sum_{l,m,n}\frac{m}{e^{(\epsilon -m \Omega)/T}-1}  \, ,
\end{equation}
and 
\begin{equation} 
e(\Omega, T) =
\sum_{l,m,n}\frac{\epsilon}{e^{(\epsilon - m \Omega)/T}-1}  \, .
\end{equation}

 In  the above equations $\Omega$  is a Lagrange multiplier whose purpose is
the implementation of angular momentum conservation [eq. (\ref{constraints})]
It has a  clear physical meaning, as the  angular velocity the system  must
counter-rotate in order to compensate black  hole angular momentum
fluctations. 

 The  total entropy  $S$  receives contributions, one   from  the  field  
entropy
\begin{equation} 
s(\Omega, T)   =  \sum_{l,m,n} \left[ (\bar{n}+1) \log (\bar{n}+1) - \bar{n} 
\log \bar{n} \right]
\,  ,
 \end {equation}                  
where   $\bar{n}$  is the mean  number of quanta sitting in  the mode 
indexed by $(l,m,n)$:    
\begin{equation} 
\bar{n}(\Omega, T) = \frac{1}{e^{(\epsilon-m \Omega)/T}-1}  \, ,
\end{equation} 
and the second from black hole entropy,   
\begin{equation} S_{\mbox{bh}}
= \frac{1}{4} 4 \pi (r_+^2 + a^2)  \, ,
\end{equation} 
where  $a=J/M$  and $r_+=  M +\sqrt{M^2-a^2}$ is the horizon radius. 

We shall be dealing with an ensemble parametrized by $\Omega$ and $T$. After
evaluating the relevant thermodynamical  calculations we want  to reproduce  
the limit $J \rightarrow 0$ which is implemented by  taking the limit $\Omega
\rightarrow 0$. Under these circunstances, the conditions for stable  thermal
equilibrium  are \footnote{The 
condition on the mixed  derivative is omitted because 
$\left(\frac{\partial^2  S}{\partial T \Omega}\right)_{\Omega=0}=0$ 
automatically.} 
\begin{eqnarray}  
\left(\frac{\partial  S}{\partial
T}\right)_{\Omega=0} &=& 0 ; \left(\frac{\partial^2  S}{\partial
T^2}\right)_{\Omega=0} <  0   \\ \left(\frac{\partial  S}{\partial
\Omega}\right)_{T, \Omega=0} &=&  0 ; \left(\frac{\partial^2  S}{\partial
\Omega^2}\right)_{T, \Omega=0} <  0    
\end{eqnarray}

Work out these conditions is much  simplified 
because $e(\Omega, T),s(\Omega, T)$ are even  functions of $\Omega$  while
$j(\Omega,T)$  is odd. Therefore,  $j(0,T)=0$, as well as the first derivatives 
$\left( \frac{\partial e}{\partial \Omega}\right)_{T \Omega=0}= \left(
\frac{\partial s}{\partial \Omega}\right)_{T \Omega=0}=0$. The computation of
these derivatives while committed  to the above constraints [eq's
(\ref{constraints})] translate the  above inequalities into:
 \begin{equation}
\left(\frac{\partial S}{\partial T}\right)_{\Omega=0} =  \left(\frac{\partial
s}{\partial T}\right)_{\Omega=0} -8 \pi M \left(\frac{\partial e}{\partial
T}\right)_{\Omega=0} =0 
\label{temp}
\end{equation}
Inspecting this  equation having an eye on the first law of thermodynamics,
we can imediatelly infer the common temperature of the system,  $T=(8\pi
M)^{-1}$. Similarly, for the second derivative: 
\begin{equation}
\left(\frac{\partial^2 S}{\partial^2 T}\right)_{\Omega=0} =
 \left(\frac{\partial^2 s}{\partial^2 T}\right)_{\Omega=0} 
- T^{-1} \left(\frac{\partial^2 e}{\partial^2 T}\right)_{\Omega=0}
 + 8 \pi \left(\frac{\partial e}{\partial T}\right)^2_{\Omega=0} \, .
\end{equation}
The first  law allows  a simplification of this expression, because
 \begin{equation}
\left(\frac{\partial^2 s}{\partial T^2}\right)_{\Omega=0} =
\frac{\partial \mbox{ } }{\partial T} \left(T^{-1} \frac{\partial e}{\partial
T}\right)_{\Omega=0} = - T^{-2} \left(\frac{\partial e }{\partial
T}\right)_{\Omega=0} + T  \left(\frac{\partial^2 e}{\partial
T^2}\right)_{\Omega=0}  
\end{equation}
and, consequently
\begin{equation}
\left(\frac{\partial^2 S}{\partial^2 T}\right)_{\Omega=0} =
 -T^{-2} \left(\frac{\partial e}{\partial T}\right)_{\Omega=0}
 + 8 \pi \left(\frac{\partial e}{\partial T}\right)^2_{\Omega=0} \, .
\label{tempf}
\end{equation}
In the thermodynamical  limit  $e = aT^4 V$ and
\begin{equation}
\left(\frac{\partial^2 S}{\partial^2 T}\right)_{\Omega=0} =
 -  4 a T V(1- 32 \pi  a T^5 V)   \, ,
\end{equation}
which is the well known result  setting a critical volume $V < V_c = (32 \pi 
a T^5)^{-1}$ for stable equilibrium \cite{dav77}.

Regarding the next two conditions, as expected $\left(\frac{\partial 
S}{\partial \Omega}\right)_{\Omega=0} =0$  is satisfied automatically. 
The only missing condition is: 
\begin{equation}
\left(\frac{\partial^2  S}{\partial^2 \Omega}\right)_{\Omega=0} = 
 \left(\frac{\partial^2  s}{\partial \Omega^2}\right)_{\Omega=0} 
- T^{-1} \left(\frac{\partial^2  e}{\partial \Omega^2}\right)_{\Omega=0} 
-\frac{2 \pi}{M^2} \left(\frac{\partial j}{\partial \Omega}\right)_{\Omega 
= 0}^2  \leq 0\, .
\label{s2o}
\end{equation}
Introducing  two new functions $\Sigma_{1,2}(T)$
\begin{eqnarray}
\Sigma_1 (T) &=& \sum_{l,m,n} m^2 \, \frac{e^{\epsilon /T}}{(e^{\epsilon
/T}-1)^2} \,\geq 0 \, ,\\
\Sigma_2(T) &=& \sum_{l,m,n} \epsilon \, m^2 \, \frac{e^{\epsilon
/T}+e^{2\epsilon /T}}{(e^{\epsilon /T} -1)^3} \,\geq 0 \, ,
\end{eqnarray}
the quantities displayed in the r.h.s of eq.(\ref{s2o}) can be
cast as:
\begin{equation} \left(\frac{\partial j}{\partial
\Omega}\right)_{\Omega=0} = T^{-1} \Sigma_1 (T)  \end {equation} \, ,
\begin{equation}
\left(\frac{\partial ^2 e}{\partial \Omega^2}\right)_{\Omega=0} = T^{-2}
\Sigma_2(T) \, ,
\end{equation}
and   finally
\begin{equation}
\left(\frac{\partial^2 s }{\partial \Omega^2}\right)_{\Omega=0}
= T^{-3} \Sigma_2(T) - T^{-2} \Sigma_1(T) \, .
\end{equation}
With  these results, eq. (\ref{s2o}) boils down  to
\begin{equation}
\left(\frac{\partial^2  S}{\partial^2 \Omega}\right)_{\Omega=0} = 
- T^{-2} \Sigma_1(T) - 2^7 \pi^3 \Sigma_1^2(T) \, ,
\label{boilled}
\end{equation}
telling us that the  system   is stable under angular velocity
fluctuations. 

We  turn next to the computation of angular momentum 
fluctuations. Because 
$\left( \frac{\partial^2 S}{\partial \Omega \partial
T}\right)_{\Omega=0}=0$, if follows that the frequency  variance 
is \cite{landau}:
\begin{equation}
(\Delta \Omega)^2 =
-\left(\frac{\partial^2  S}{\partial^2 \Omega}\right)_{\Omega=0}^{-1}
\end{equation} 
which yields for the angular momentum fluctuations:
\begin{equation}
(\Delta  J)^2 = \left(\frac{\partial j}{\partial
\Omega}\right)_{\Omega=0}^2 (\Delta \Omega)^2  = 
\frac{\Sigma_1(T)}{ 1 + 2^7 \pi^3 T^2 \Sigma_1(T)} \, .
\end{equation}

Turning now to the computation of $\Sigma_1$, we recall  the asymptotic
expansion  of the  spherical Bessel function 
\begin{equation}
j_l(x) \approx x^{-1}  \cos (x -(l+\frac{1}{2})
\frac{\pi}{2} - \frac{\pi}{4} )
\end{equation}
Therefore, the roots are approximately localized  at
\begin{equation}
x_{l n} \approx (\frac{l}{2} + n) \pi
\end{equation}
Indexing the  modes by a new triple of quantum numbers  $m,l$
and $s=l/2+n$ ($ -l \leq m \leq l, l \leq 2s$) is a convenient expedient
because $\epsilon=\epsilon_s=s \pi/R$ and the  modes become degenerate  with
respect to the   the remaining quantum  numbers $m,l$. With the  aid of the
identity:  \begin{equation} 
\sum_{l=0}^{2s} \sum_{m=-l}^{l} m^2= \frac{2}{3} s(s+1)(2s+1)^2  \, ,
\end{equation}
we can reexpress 
\begin{equation}
\Sigma_1(T)  = \frac{2}{3}\sum_s s(s+1)(2s+1)^2
\frac{e^{\epsilon_s/T}}{(e^{\epsilon_s/T}-1)2} \, .
\end{equation}

Next, we go to the thermodynamical limit $R  \rightarrow \infty$. Introducing 
the auxilliary  quantities $x=\epsilon/T$  and $\alpha = R T/\pi$,
\begin{equation}
\Sigma_1 \approx \frac{2}{3} \alpha^2 \int_0^{\infty} x(\alpha x +1)(2\alpha
x+1)^2 \frac{e^x}{(e^x-1)^2} dx \, .
\end{equation}
Assumming that the  vessels'  radius is  always much larger than the black
hole, which is also equivalent to a thermal wave length much larger
than the size of the cavity  ($\alpha >>1$), then, the leading  term in this
expression is 
\begin{equation} \Sigma_1(T) \approx \frac{8}{3}
\frac{R^5T^5}{\pi^3} \left[\frac{1}{\pi^2} \int_0^\infty\ x^4
\frac{e^x}{(e^x-1)^2} dx\right] \, . \end{equation} Because the function 
\begin{equation}
a(\lambda)= \frac{1}{\pi^2} \int_0^\infty \frac{x^3}{e^{\lambda x}-1}
\end{equation}
is such that $-a'(1)$  reproduces the bracket in the previous equation and 
$a(\lambda) = a(1) \lambda^{-4}$, where $a(1)=\pi^2/60$ is the
Stephan-Boltzmann's constant:
\begin{equation}
\Sigma_1(T) \approx  \frac{8}{45\pi} R^5 T^5  \, .
\label{s1}
\end{equation}
Thus
\begin{equation}
(\Delta  J)^2 = \frac{8}{45\pi} \frac{R^5 T^5}{1+2^{10} \pi^2/45 R^5 T^7} \, .
\label{dj}
\end{equation}
 For a spherical cavity the bound on
the critical  volume translates  into $T^2 \leq (45/32 \pi^4) (RT)^{-3}$,
which  in the limit $R/2M \gg 1$ yields  a lower bound on the these
fluctuations: 
\begin{equation}
(\Delta  J)^2 \stackrel{>}{\sim} \frac{2 \pi^2}{45} R^3 T^3.
\end{equation}

	This shows that the fluctuations are  governed by the size of the
 cavity.

\section{charge fluctuations}

In this section  we shall deal with  charge fluctuations. For this
end we  consider   a vessel of volume V containing  scalar charged 
quanta in  equilibrium with a black  hole. Similarly to the previous procedure,
we assume  that the black hole has a  nonvanishing  charge, consider the
equilibrium conditions and  only at the end take  the limit where the charge  
does vanish. Energy  and charge bookeeping requires 
\begin{eqnarray} 
E &=& \eta(\mu, T) V + M \\ 0 &=& \rho(\mu,T) + Q  \, , \end{eqnarray}
were $Q$  and $M$   are the black hole's  mass and charge; $\rho(\mu,T)$
 and $\eta(\mu, T)$   are the charge and energy densities  of the field,
respectively: 
\begin{equation}
\rho(\mu,T) = \frac{q}{(2\pi)^3} \left[ \int \frac{d^3k}{e^{(\epsilon-\mu)/T}-1}
-\int \frac{d^3k}{e^{(\epsilon+\mu)/T}-1}\right] \, ,
\end{equation}
\begin{equation}
\eta(\mu,T) = \frac{1}{(2\pi)^3} \left[ \int
\frac{\epsilon d^3k}{e^{(\epsilon-\mu)/T}-1} +\int
\frac{\epsilon d^3k}{e^{(\epsilon+\mu)/T}+1}\right] \, ,
\end{equation}
and $q$  is the elementary charge of the field. We assumed that the total
system is neutral. Here, the chemical potential $\mu$  implements
the charge conservation constraint and has the physical meaning  of an
eletrostatic energy associated to the complete screening the black  hole
charge. 

The  entropy densities of the species are 
\begin{equation}
s_{\pm} = \int
 \frac{d^3k}{(2\pi)^3} \left[((\bar{n}_{\pm}+1) \log (\bar{n}_{\pm}+1)-
\bar{n}_{\pm} \log \bar{n}_{\pm} \right] \, ,
\label{s+-}
\end{equation}
with
\begin{equation}
\bar{n}_{\pm} =  \frac{1}{e^{(\epsilon \mp \mu)/T}-1} \, .
\label{mn}
\end{equation}
  The system's entropy is 
\begin{equation}
S= (s_+ + s_-) V + \frac{1}{4} 4 \pi r_+^2 
\end{equation}
were $r_+ = M + \sqrt{M^2-Q^2}$

The ensemble is now  parametrized by $\mu$ and  $T$.Because
$Q$  is an odd function of $\mu$, the limit $Q  \rightarrow 0$  which  we  are
interested in is reproduced by   $\mu \rightarrow 0$.

It is a  trivial  exercise to  compute:
\begin{equation}
\left( \frac{\partial  S}{\partial T}\right)_{\mu=0} = 8 a T^2 V (1-8 \pi M T)
\end{equation} 
Whose vanishing reproduces the black  hole  temperature $T^{-1} = 8\pi M$.
The second derivative provides the critical volume :
\begin{equation}
\left( \frac{\partial^2  S}{\partial T^2}\right)_{\mu=0} = -8 a T V (1 - 64
\pi V a T^5) \, ,
\end{equation}
which is  half  of the previous one, because it is
inversely  proportional to the number of species. 
Likewise, the first derivative $\left( \frac{\partial  S}{\partial
\mu }\right)_{\mu=0}=0$ vanishes automaticaly and does not provide any  new
condition. Again, the calculation of the second derivative is much 
simplified by the fact that the first derivatives of the field entropy and
energy  with respect to $\mu$ vanish for $\mu=0$. Calling $s=s_+ + s_-$, 
\begin{equation} \left(
\frac{\partial^2  S}{\partial \mu^2}\right)_{\mu=0}=  V \left( \frac{\partial^2 
s}{\partial \mu^2}\right)_{\mu=0} - \frac{V}{T} \left( \frac{\partial^2 
\eta}{\partial \mu^2}\right)_{\mu=0} -4\pi V^2 \left( \frac{\partial
\rho}{\partial \mu}\right)_{\mu=0}^2 
\end{equation}
Calling $x_\pm=(\epsilon \mp\mu)/T$, by virtue of eqs.(\ref{s+-})  and 
(\ref{mn})
\begin{equation}
\left(\frac{\partial s_\pm}{\partial x}\right) =   \int \frac{d^3k}{(2\pi)^3}x
\frac{\partial n}{\partial x} \, .
\end{equation}
Thus,
\begin{eqnarray}
\left(\frac{\partial^2  s}{\partial \mu^2}\right)_{\mu=0} &=& 2 T \int
\frac{d^3x}{(2\pi)^3} \left[\frac{\partial n}{\partial x} + x \frac{\partial^2
n}{\partial x^2}\right] \\
\left(\frac{\partial^2  \eta}{\partial \mu^2}\right)_{\mu=0} &=& 2 T^2  \int
\frac{d^3x}{(2\pi)^3} x  \frac{\partial^2 n}{\partial x^2}\\  
\left(\frac{\partial \rho}{\partial \mu}\right)_{\mu=0} &=& -2 q T^2 \int
\frac{d^3x}{(2\pi)^3} \frac{\partial n}{\partial x}
\end{eqnarray}
Having in mind that 
\begin{equation}
\int \frac{d^3x}{(2\pi)^3} \frac{\partial n}{\partial x} =
-\int \frac{d^3x}{(2\pi)^3} \frac{e^x}{(e^x-1)^2} =-\frac{1}{6}
\end{equation} 
and putting all these pieces together:
\begin{equation}
\left(\frac{\partial^2  S}{\partial T^2}\right)_{\mu=0}= -
\frac{1}{9} T V \left(3 + 4 \pi V q^2 T^3 \right) \leq 0 \, ,
\end{equation}
which  ensures stability under eletrostatic energy fluctations. Paralleling the
calculation of angular momentum fluctuations we estimate charge fluctuations 
as follows:
\begin{equation} 
\Delta Q^2 = -\left(V \frac{\partial \rho}{\partial
\mu}\right)_{\mu=0}^2 \left[\left(\frac{\partial^2  S}{\partial
\mu^2}\right)_{\mu=0}\right]^{-1} \approx \frac{q^2 V T^3}{\left(3+ 4 \pi V
q^2 T^3 \right)} \, .  
\end{equation}
Because the reservoir is always  much larger than  the hole ($T^3 V \gg 1$), it
follows that
\begin{equation}
\frac{\Delta Q^2}{\hbar c} \approx \frac{1}{4\pi}  \, ,
\end{equation}
independently of the size of the cavity.

\section{assessements}

We  can sum up our resuts by telling  that Schwarzschild black holes undergo a
considerable amount of fluctuations both in angular momentum and charge which 
should not be neglected.  Because the mean angular momenta squared  of the
particles in the thermal bath $\bar{J^2} \sim \bar{p^2 b^2}$ where $b$
is the impact parameter, one would  expect the angular momentum spread
to  scale roughly linearly with the size  of the cavity. We obtained that it
actually scales  with  power $3/2$, not far from our intuitive feeling.
Thus, we can regard  black   hole angular momentum fluctuations
to be the outcome of the  random absorption and emission of quanta with  random 
angular momenta from the thermal bath. 

Furthermore, if charge fluctuations were the outcome  of  
random absorption and emission of charged  quanta, then one would expect the
black  hole charge spread  to  be independent on the size of the cavity.
As  a matter of fact,  in the  limit $R/2M \gg 1$  we obtained that
$\Delta Q^2/(\hbar c) =1/(4\pi)$ lending support to this interpretation.
Nevertheless, there is a hidden puzzle here. If we insist that this is the
appropriate physical description of the spread then this spread had to be
proportional to the elementary  charge of the field, which our  calculations
showed not to   be the case. This poses  a serious dilemma, because either
black hole physics sets atomatically the scale of the elementary  charge to be
such  that  $\alpha \sim 1/4\pi$ (this yields an elementary  charge  which is
only  three times the charge of the electron), or the  physical process behind
these fluctuations is a mistery  which  remains to be cleared up.

 A major advantage of dealing with black holes   in termal contact with
radiation  is that we can  gloss over the issue  of the hole's absorptivity
$\Gamma(\omega,m,q)$, whose computation  is in general very   cumbersome. The
reason is that  after many and many reflections against a grey 
body (  black holes are grey bodies \cite{bek}-\cite{mar}) inside the
cavity, the outgoing radiation becomes truly black body, in which case
$\Gamma_{\mbox{\small{effective}}}=1$. This brings about a tremendous
simplification of the problem  but has its  price, we cannot scrutinize black 
hole fluctuations  at Planckian  scales because the stability  condition  
combined with  the fact  that $2M < < R$  imposes that $T >>T_p$. If these
fluctuations survive at these large energy  scales, is   currently
under investigation by a very different strategy.

\section*{acknowledgements} I am  indebted to Fapesp  for financial support and
to J. Bekenstein for many enlightening conversations.

\end{document}